\documentclass[structabstract]{aa}
\usepackage{txfonts}
\usepackage[utf8]{inputenc}
\usepackage[T1]{fontenc}
\usepackage{natbib}
\usepackage[breaklinks=true]{hyperref}
\usepackage{amssymb}
\usepackage{graphicx}
\usepackage{breakurl}
\renewcommand{\arraystretch}{1.2}

\bibpunct{(}{)}{;}{a}{}{,} 
\title{The average GeV-band Emission from Gamma-Ray Bursts}
\author{J. Lange\inst{1} \and M. Pohl\inst{1,2}}
\institute{DESY, 15738 Zeuthen, Germany \and Institute of Physics and Astronomy, University of Potsdam, 14476 Potsdam, Germany}
\date{Received / Accepted}
\abstract {} {We analyze the emission in the $0.3$ -- $30 \ \mathrm{GeV}$ energy range of Gamma-Ray Bursts detected with the \textit{Fermi} Gamma-ray Space Telescope. We concentrate on bursts that were previously only detected with the Gamma-Ray Burst Monitor in the $\mathrm{keV}$ energy range. These bursts will then be compared to the bursts that were individually detected with the Large Area Telescope at higher energies.} {To estimate the emission of faint GRBs we use non-standard analysis methods and sum over many GRBs to find an average signal which is significantly above background level. We use a subsample of $99$ GRBs listed in the Burst Catalog from the first two years of observation.} {Although mostly not individually detectable,
the bursts not detected by the Large Area Telescope on average emit a significant flux in the energy range from $0.3 \ \mathrm{GeV}$ to $30 \ \mathrm{GeV}$, but their cumulative energy fluence is only 8\% of that of all GRBs. Likewise, the GeV-to-MeV flux ratio is less and the GeV-band spectra are softer. We confirm that the GeV-band emission lasts much longer than the emission found in the $\mathrm{keV}$ energy range. The average allsky energy flux from GRBs in the GeV band is $6.4\cdot 10^{-4}\ {\rm erg\,cm^{-2}\,yr^{-1}}$ or only $\sim 4 \%$ of the energy flux of cosmic rays above the ankle at $10^{18.6}$~eV.} {}

\begin{document}

\titlerunning{GeV-band Emission from GRBs}
\authorrunning{J. Lange \& M. Pohl}
\maketitle

\section{Introduction}

Since its launch in June 2008 the \textit{Fermi} Gamma-ray Space Telescope has broadened our understanding of Gamma-Ray Bursts (GRBs). The two main instruments on board of the satellite are the Gamma-Ray Burst Monitor (GBM, \citet{2009ApJ...702..791M}) and the Large Area Telescope (LAT, \citet{2009ApJ...697.1071A}). Together they are capable of observing GRBs over 7 decades of energy. Especially the LAT that covers the high-energy region from $30\ \mathrm{MeV}$ to $300 \ \mathrm{GeV}$ has the potential to provide new insight into the underlying physics of GRBs. However only a small fraction of the GRBs detected with the GBM were individually detected with the LAT \citep{2009ARA&A..47..567G}. 

Whereas the keV-MeV emission may well be quasi-thermal
emission, i.e. of photospheric origin \citep{2000ApJ...529..146E,2009ApJ...702.1211R,2012MNRAS.420..468P}, the GeV-band emission is indicative of particle acceleration to very high energies and can potentially probe whether or not GRBs
are powerful enough to provide a significant part of ultrahigh-energy cosmic rays 
\citep{2011ARA&A..49..119K,2011ApJ...738L..21E}.
Recently, the GeV-band photon output of GRBs was estimated on the basis of the relatively few GRBs, that are
individually detected with \textit{Fermi}-LAT, and {was found to be small} in comparison
with the source power needed to sustain cosmic rays above the ankle in the spectrum, at $10^{18.6}$~eV 
\citep{2010ApJ...722..543E}. One source of uncertainty in this statement is the unknown level of high-energy
photon output of the many GRBs not individually detected with LAT \citep{2010arXiv1010.5007W}.

Here we re-analyze data of the \textit{Fermi}-LAT detector with a view to infer the GeV-band high-energy emission from GRBs. In earlier studies the focus had been placed on studying each burst individually 
\citep[e.g.][]{2011ApJ...734L..27A,2012ApJ...754..121T,2012ApJ...745...72Z,2012ApJ...756...64Z}. However, by analyzing many GRBs together one is able to obtain more precise results for the emission of GRBs. Therefore, we placed the focus on bursts that do not show a significant signal in the $\mathrm{GeV}$ range when analyzed individually. A subsample of $99$ GRBs has been defined that were in principle detectable with \textit{Fermi}-LAT and occurred in regions of low background emission coming from other galactic and cosmic sources. For each burst the spectrum of expected background photons and an effective area is estimated. Together with the spectrum of observed photons we thus determine the fluence from these bursts.

\section{Framework}

Most of the GRBs detected by the GBM do not trigger the LAT. Therefore it is expected that most of the GRBs do not show a significant signal above background. The bursts that did not trigger the LAT will be referred to as GBM-detected bursts. We determine their emission by counting LAT-detected photons within a certain time interval and a certain solid-angle element, henceforth referred to as the Area of Integration (AOI). The choice of the time window of observation and the AOI is important, because
GeV emission can be delayed \citep{2010ApJ...712..558A} and the GRB position is typically not well determined by the GBM. For long observation times and a large AOI the signal might not be distinguishable from the background while for short observations and a small AOI the emission could simply be missed. A few general considerations are therefore in order.

\subsection{Energy Intervals}

In general the LAT is designed to detect photons in an energy range of $30 \ \mathrm{MeV}$ to $300 \ \mathrm{GeV}$ \citep{2009arXiv0907.0626R}. The analysis could in principle be performed for the entire energy range, but the sensitivity of the detector varies with energy. The effective area of the LAT has been derived through Monte Carlo simulations and {verified with flight data} \citep{2012ApJS..203....4A}, and it is is very low for energies around 100~$\mathrm{MeV}$ in comparison to $\mathrm{GeV}$ energies. Thus at low energies the number of detected photons is low, but they represent a large portion of the flux because the actual photon flux is normally a decreasing function of the energy \citep{2012arXiv1202.4039T}. Additionally, the reconstruction of the photon arrival direction is less precise for lower energies, which further reduces the signal-to-background ratio. As a result the detection significance for energies below $\lesssim 300 \ \mathrm{MeV}$ should be very low and the flux determination uncertain. We therefore analyze the energy interval from $300 \ \mathrm{MeV}$ to $30 \ \mathrm{GeV}$ only.

\subsection{Time Intervals}

Recent studies of the $\mathrm{GeV}$-band emission from GRBs suggest a significant production of highly energetic photons long after the prompt emission at keV--MeV energies \citep[e.g.][]{2009arXiv0907.0714T, 2012MNRAS.421L..14R}, mandating that the GRBs be monitored for long time intervals after the prompt emission phase. The duration of GRBs is characterized by $T_{90}$, the mid time in which 
$90 \%$ of the fluence is observed in the BATSE energy range ($50 \ \mathrm{keV}$ to $300 \ \mathrm{keV}$). The duration of the GBM emission phase and the LAT emission phase is likely to be correlated and therefore it is reasonable to choose time intervals as multiples of $T_{90}$ rather than absolute time periods.

\subsection{Area of Integration}

\begin{figure}[ht]
 \resizebox{\hsize}{!}{\includegraphics[angle=0]{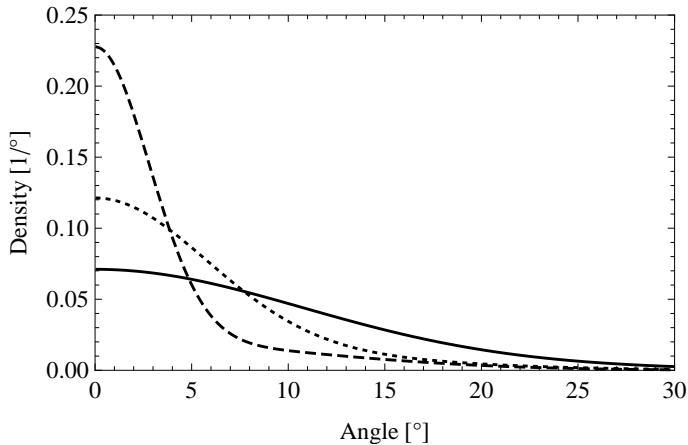}}
 \caption{The expected GBM localization accuracy for $1^\circ$ (dashed), $5^\circ$ (fine dashed) and $10^\circ$ (continous) statistical error radius. A normal distribution is assumed for the statistical error.}
 \label{GBMDistribution}
\end{figure}

From the GBM and other observatories the position of every burst is known with a precision of up to $1 ^{\circ}$ while data from other observatories or a combined analysis may permit a localization in the arcsecond range. A comparison of such localizations indicates a systematic error in the positioning of the GRBs with GBM. The best fit for the systematic error is the combination of 2 Gaussians with dispersion $2.6^\circ$ with $72\%$ weight and $10.4^\circ$ with $28\%$ weight, to be added in quadrature to the statistical error \citep{2012ApJS..199...18P}. The expected distribution of GBM burst localization errors are shown in Fig. \ref{GBMDistribution}. For all other localization sources we neglect systematic errors.

The imperfect angular reconstruction of the LAT, characterized by the point-spread function (PSF), may lead to a loss of events, even if the actual GRB was located within the AOI. Therefore the radius of the AOI will be the statistical error added in quadrature with $2^\circ$ (PSF, $68\%$ containment for $300 \ \mathrm{MeV}$) and $3^\circ$ (systematic error). In the case of GRB positioning with other detectors than GBM the systematic uncertainty is set to $0^\circ$. As a figure of merit roughly $\sim 2/3$ of all observed GeV photons of the GBM-detected bursts are expected to lie within the angular radius

\begin{equation}
 \alpha_{\rm AOI} = \sqrt{\sigma_{\rm stat}^2 + \sigma_{\rm sys}^2 + \sigma_{\rm PSF}^2}\ .
 \label{AOI}
\end{equation}

As we sum the number of gamma-ray events over many GRBs, the global photon-detection efficiency is 
approximately 2/3.

\section{Sample Definition}

For the analysis we use a sample of GRBs detected by the GBM from the beginning of normal science operation of the \textit{Fermi} satellite on August 04, 2008 until July 9, 2010 as listed in \citet{2012ApJS..199...18P}. In this time interval $472$ bursts triggered the GBM. Some of these GRBs have to be excluded because they did not lie in the Field of View (FoV) of the LAT or were not observed during normal science operation of {\it Fermi}. It is also useful to exclude certain GRBs that are expected to have a low signal to background ratio. After applying all cuts $99$ bursts remain. 14 of them have been previously detected by the LAT.

\subsection{LAT Field of View}

The GBM can detect GRBs in almost every direction. On the other hand the LAT has a FoV of $\sim 2.4 \ \mathrm{sr}$ at $1 \ \mathrm{GeV}$ after all analysis cuts for background rejection have been made \citep{2009ApJ...697.1071A}. This means that only some of the GRBs detected by the GBM are actually in the FoV of the LAT and therefore detectable by it. The sensitivity in terms of the effective area is a decreasing function of the angle $\theta$ between the photon arrival direction and the LAT boresight. Photons arriving at $\theta \gtrsim 75 ^\circ$ likely escape detection. It is therefore useful to concentrate only on GRBs where the angle between the GRB position and the LAT boresight during the observation was $\le 70 ^\circ$.

\subsection{Data Quality}
\label{Data Quality}

Some GRBs have to be rejected because of the quality of the data even if they are in the FoV of the LAT. As recommended by the LAT team, {\it Fermi} is required to be outside the South Atlantic Anomaly (SAA) and in normal science data-taking mode during the observation time. Additionally, an Earth-relative zenith-angle cut of $100^\circ$ is imposed to prevent the spill-in of emission from the Earth's limb. All bursts are required to fulfill this criterion at least in a time interval of $10 \times T_{90}$ starting from the beginning of the $T_{90}$ time.

\subsection{Background Emission}

To avoid a large number of background photons it is reasonable to exclude certain regions of sky that are known to have a particularly high background intensity in the high-energy region. The dominant background sources are diffuse emission from the galactic plane as well as bright gamma-ray sources like pulsars. It has been found that the photon intensity at GeV energies increases by roughly two orders of magnitude when observing the galactic plane \citep{2012arXiv1202.4039T}. Regions with a galactic latitude $b$ of $|b| < 5^{\circ} + \alpha_{\rm AOI}$ will be excluded, where $\alpha_{\rm AOI}$ is the radius of the AOI specified in equation \ref{AOI}. It is also useful to exclude regions around very bright gamma-ray sources. The most luminous ones are the Blazar 3C454.3, the Vela Pulsar, the Crab Pulsar, the Geminga Pulsar, and the pulsars PSR J1709-4429 and PSR J2021+4026. The angular separation to these sources is required to be at least $1^\circ + \alpha_{\rm AOI}$. Except for the Blazar 3C454.3 all these sources lie within the galactic plane and have already been excluded.

\subsection{Localization Error}

Finally it is also useful to exclude GRBs with a high localization error because a large AOI with correspondingly high total background would have to be considered. By requiring that the statistical error radius be smaller than $5^\circ$ the number of samples is decreased by a factor of $\sim 2/3$. There is a correlation between the statistical error radius and the fluence. Therefore, by applying this cut on the error angle mostly faint bursts are rejected.

\section{Method}

\subsection{Modeling the Background}

The {\it Fermi} Science Tools are used for the analysis of the bursts. Although for the duration of a single GRB the background emission is almost negligible, it becomes crucial when summing over many GRBs. The main sources of background radiation are the galactic and extragalactic diffuse emission, misclassified events, and individual sources like pulsars. 

Using binned likelihood analysis we construct a model of instrumental background and the gamma-ray sky by fitting data obtained in a time interval of $3 \times 10^6 \ \mathrm{s}$ until $1$ hour before the trigger time of each burst. The pointing history of the spacecraft during the emission phase of the GRB then permits a prediction of the expected background for each burst.
The event files and the spacecraft files are available at the \textit{Fermi} mission website \footnote{\url{http://heasarc.gsfc.nasa.gov/}}. The LAT has different instrument response functions (IRFs). The IRFs differ especially in the efficiency in terms of the effective area, the PSF and the energy dispersion. As recommended by the LAT team the P7SOURCE\_V6 class is used together with the corresponding galactic diffuse model (gal\_2yearp7v6\_v0) and isotropic spectral template (iso\_p7v6source). It offers very high data 
quality with low background from misclassified photons and a fairly large effective area.\par\medskip

The Region of Interest (ROI) for modeling the background emission is a circle around the burst location with a radius of $22^\circ$; the energy range is $300 \ \mathrm{MeV}$ to $30 \ \mathrm{GeV}$. For the analysis only time intervals are used during which the LAT was not in the SAA, in normal science-operation mode, and in the normal range of rocking angles of not more than $52^\circ$. Additionally the ROI is required to not overlap with the Earth's limb. The point and extended sources are listed in the \textit{Fermi} LAT Second Source Catalog which is based on two years of observations \citep{2012ApJS..199...31N}. As recommended by the LAT team ten energy bins per decade and angular pixels of size $0.2$ degrees are used. A first fit is performed with the DRMNGB optimizer and the result is then used for a second fit with the NEWMINUIT optimizer to get more precise results.

\subsection{Analysis of the Emission Phase}

For the analysis of the emission phase the P7SOURCE\_V6 class is also used. The number of expected background photons is calculated by computing an exposure map for the AOI. In this case the exposure map describes how each source at the sky contributes to the observed photon {counts} inside the AOI. It is defined as:

\begin{equation}
 \epsilon(E, \hat{p}) = \int_{\rm AOI} dt \ d\hat{p}_{\rm obs} \ R(E ,\hat{p}, \hat{p}_{\rm obs}) \ \mbox{.}
\end{equation}

$E$ is the energy, $\hat{p}$ and $\hat{p}_{obs}$ the true and observed direction and $R$ is the response derived from the IRFs.  This can be used to estimate the background photons inside the AOI from the background model. Energy dispersion has been neglected in this step. To get an estimate on the fluence of the GRB the exposure has to be estimated. The possible GRB positions are distributed throughout the AOI. The best estimate for the exposure of the GRB is the integral of the exposure convolved with the distribution function $\phi$ of the expected localization errors (see for example Fig. \ref{GBMDistribution}).

\begin{equation}
 \epsilon(E) = \int_{\rm Sky} d\hat{p} \ \epsilon(E, \hat{p}) \ \frac{\phi(\delta\hat{p})}{2\pi\sin(\delta\hat{p})}
\end{equation}

In this way the error of the burst localization and the fact that some GRBs actually lie outside the AOI is automatically taken into account. 

\section{Results}

It is useful to first look at the raw photon counts in comparison to the number of expected photons for GRBs not detected by the LAT. The results are shown in Table \ref{Photons_Temporal}.

\begin{table}[ht]
 \centering
 \begin{tabular}{c c c c c}
  \hline
  $T_{\mathrm{start}} [T_{90}]$ & $T_{\mathrm{stop}} [T_{90}]$ & GRBs & $n_{\rm obs}$ &$\lambda$ \\
  \hline
  $-5$ & $-2.5$ & $75$ & $4$ & $2.35$ \\
  $-2.5$ & $0$ & $79$ & $1$ & $2.90$ \\
  $0$ & $2.5$ & $85$ & $13$ & $3.53$ \\
  $2.5$ & $5$ & $85$ & $7$ & $3.68$ \\
  $5$ & $7.5$ & $85$ & $8$ & $3.83$ \\
  $7.5$ & $10$ & $85$ & $11$ & $3.93$ \\
  $10$ & $12.5$ & $79$ & $5$ & $3.26$ \\
  $12.5$ & $15$ & $76$ & $5$ & $2.99$ \\
  $15$ & $17.5$ & $76$ & $6$ & $3.02$ \\
  $17.5$ & $20$ & $73$ & $3$ & $2.92$ \\
  $20$ & $22.5$ & $71$ & $8$ & $2.36$ \\
  $22.5$ & $25$ & $67$ & $4$ & $2.17$ \\
  $25$ & $27.5$ & $64$ & $3$ & $1.92$ \\
  $27.5$ & $30$ & $64$ & $5$ & $1.88$ \\
  $30$ & $32.5$ & $63$ & $1$ & $1.78$ \\
  $32.5$ & $35$ & $61$ & $1$ & $1.57$ \\
  \hline
  $0$ & $10$ & $85$ & $39$ & $14.97$ \\
  \hline
 \end{tabular}
\caption{The number of {observed photons ($n_{\rm obs}$) and the expected background ($\lambda$)}
for the GBM-detected bursts for the different time intervals. The times are given in multiples of the $T_{90}$ time of each burst and relative to the start of the $T_{90}$ time interval. For times outside the $0$ -- $10$ $T_{90}$ time interval some GRBs have to be excluded on account of cuts that were originally only applied in this time interval.}
\label{Photons_Temporal}
\end{table}

As we can see there is a significant emission in the time interval from $0$ to $10$ times the $T_{90}$ time. In this time $\lambda=14.97$ photons are expected and $n=39$ observed. The chance probability for a background fluctuation is $1.7 \times 10^{-7}$, based on Poisson statistics

\begin{equation}
 P(n \geq n_{obs},\lambda) = \sum \limits_{n=n_{\rm obs}}^{\infty} \frac{\lambda^n}{n!} e^{-\lambda}  \ .
\label{poisson}
\end{equation}

This is a significant signal when summed over all GRBs but clearly not significant for a single burst for which the number of observed counts is typically smaller than $1$ and the number of observed photons mostly $0$ or $1$. For the $11$ observed short bursts ($T_{90} < 2 \ \mathrm{s}$) there was $1$ observed photon and $0.06$ expected background photons which is too low a signal to claim a detection. For the remaining $74$ long bursts ($T_{90} > 2 \ \mathrm{s}$) there were $38$ observed photons and $14.91$ expected from background. Note that before the GRB trigger we observe 5 events and expect 5.25. Later than $25\,T_{90}$ after trigger we observe 10 events and expect 7.15, indicating that there is little, if any, emission at that late stage, and no evidence for any activity before the GBM trigger.

\subsection{Detection Significance}

One can search for significant signals of single GRBs in these data. For this purpose the number of observed and predicted photons are compared for the 4 time intervals between 0 to 10~$T_{90}$ as well as the total time interval. As a cross-check the LAT-detected bursts will also be tested with the same method. $99$ bursts are tested for 5 different time intervals and therefore we have at most $495$ trials, because the time intervals are not all independent. Therefore the probability that the background emission, $\lambda$, reaches or exceeds the observed emission, $n_{\rm obs}$, by normal fluctuations should be smaller than $10^{-4}$ for a 5\% post-trials chance probability.
All $14$ bursts previously detected by the LAT show a significant signal in at least one time interval. This is not surprising as these bursts actually triggered the LAT and also have high-precision localizations either by the LAT or other observatories. However there were $2$ other bursts with significant emission: GRB 100207B and GRB 081009A. The latter one has also been reported as a candidate for a LAT-detected burst with the same method but with a wider energy range ($E > 100 \ \mathrm{MeV}$) and longer observations times ($t_{\rm obs} = 1500 \ \mathrm{s}$) \citep{2012MNRAS.421L..14R}. However it has not been added to the LAT Burst catalog so far. The results of this analysis are shown in Table \ref{DetectionSignificance}.

\begin{table}[ht]
 \centering
 \begin{tabular}{c c c c c c}
  \hline
  GRB & Time $[T_{90}]$ & $n_{\rm obs}$ & $\lambda$ & $P(n \geq n_{\rm obs},\lambda)$\\
  \hline
  081009A & $5$ -- $7.5$ & $3$ & $0.0366$ & $7.93 \times 10^{-6}$ \\
  100207B & $0$ -- $2.5$ & $2$ & $0.0123$ & $7.54 \times 10^{-5}$ \\
  \hline
 \end{tabular}
\caption{{We show results for} the two GRBs not listed in the LAT Burst catalog that show significant emission in the analysis. As before, $n_{\rm obs}$ is the number of observed photons and $\lambda$ the expected number of photons from background.}
\label{DetectionSignificance}
\end{table}

\subsection{Photon Fluence}

\begin{figure}[ht]
 \resizebox{\hsize}{!}{\includegraphics[angle=0]{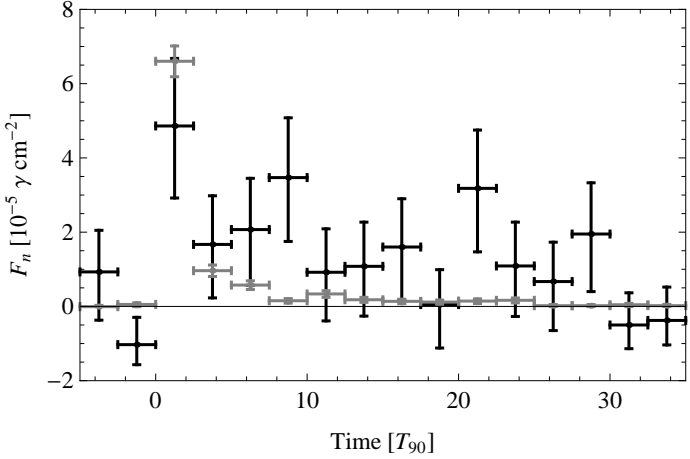}}
 \caption{Average fluences for different time intervals in units of $T_{90}$. Shown in black (grey) are the average fluences for the GBM-detected (LAT-detected) bursts. The fluences for the LAT detected bursts were scaled by a factor of $10^{-2}$ to fit on the same plot.}
 \label{Lightcurve}
\end{figure}

The average photon fluence of GRBs can be calculated by separately summing the exposure and the observed photon and expected background counts in every energy bin. {To estimate the cumulative exposure over the entire energy range between 300~MeV and 30~GeV we assume the emission
follows a power-law with index of $-2.3$. As we shall see below, this index is appropriate for LAT-detected GRBs,
whereas the average emission of GBM-detected bursts is better described with a power-law index
of $-3$, in which case the photon fluence would be underestimated by about 10\%. The $68 \%$-confidence
statistical uncertainty of the fluence has been calculated using an incomplete gamma function, i.e. the Bayesian inversion of a Poisson distribution for uniform prior, and is considerably larger than the systematic uncertainty in estimating the fluence.} The analysis has been done for the GBM- and LAT-detected bursts separately. The result is shown in Table \ref{PhotonFluence_Temporal} and
graphically displayed in Figure~\ref{Lightcurve}.

\renewcommand{\arraystretch}{1.5}
\begin{table}[ht]
 \centering
 \begin{tabular}{c c c}
  \hline
  Time $[T_{90}]$ & \multicolumn{2}{c}{Fluence [$\gamma \ \mathrm{cm}^{-2}$]} \\ 
   & GBM & LAT \\ 
  \hline
  $(-5)$ -- $(-2.5)$ & $0.93_{-1.30}^{+1.12} \times 10^{-5}$ & $-0.64^{+3.47} \times 10^{-5}$ \\
  $(-2.5)$ -- $0$ & $-10.29_{-5.40}^{+7.35} \times 10^{-6}$ & $5.34_{-5.98}^{+4.55} \times 10^{-5}$ \\
  $0$ -- $2.5$ & $4.86_{-1.94}^{+1.82} \times 10^{-5}$ & $66.03_{-4.17}^{+4.12} \times 10^{-4}$ \\
  $2.5$ -- $5$ & $1.67_{-1.44}^{+1.31} \times 10^{-5}$ & $9.63_{-1.57}^{+1.52} \times 10^{-4}$ \\
  $5$ -- $7.5$ & $2.07_{-1.50}^{+1.38} \times 10^{-5}$ & $5.75_{-1.19}^{+1.15} \times 10^{-4}$ \\
  $7.5$ -- $10$ & $3.47_{-1.72}^{+1.61} \times 10^{-5}$ & $15.34_{-6.48}^{+5.92} \times 10^{-5}$ \\
  $10$ -- $12.5$ & $0.92_{-1.31}^{+1.17} \times 10^{-5}$ & $33.64_{-9.19}^{+8.69} \times 10^{-5}$ \\
  $12.5$ -- $15$ & $1.08_{-1.34}^{+1.19} \times 10^{-5}$ & $18.00_{-7.00}^{+6.44} \times 10^{-5}$ \\
  $15$ -- $17.5$ & $1.60_{-1.44}^{+1.30} \times 10^{-5}$ & $13.71_{-6.33}^{+5.71} \times 10^{-5}$ \\
  $17.5$ -- $20$ & $0.05_{-1.17}^{+0.94} \times 10^{-5}$ & $11.75_{-6.05}^{+5.37} \times 10^{-5}$ \\
  $20$ -- $22.5$ & $3.18_{-1.71}^{+1.57} \times 10^{-5}$ & $14.42_{-6.62}^{+5.98} \times 10^{-5}$ \\
  $22.5$ -- $25$ & $1.09_{-1.36}^{+1.18} \times 10^{-5}$ & $16.49_{-6.90}^{+6.30} \times 10^{-5}$ \\
  $25$ -- $27.5$ & $0.67_{-1.32}^{+1.06} \times 10^{-5}$ & $1.96_{-2.37}^{+3.23} \times 10^{-5}$ \\
  $27.5$ -- $30$ & $1.95_{-1.55}^{+1.38} \times 10^{-5}$ & $1.95_{-2.36}^{+3.20} \times 10^{-5}$ \\
  $30$ -- $32.5$ & $-5.00_{-6.37}^{+8.66} \times 10^{-6}$ & $4.48_{-4.90}^{+3.72} \times 10^{-5}$ \\
  $32.5$ -- $35$ & $-3.77_{-6.60}^{+8.98} \times 10^{-6}$ & $2.06_{-2.46}^{+3.35} \times 10^{-5}$ \\
  \hline
 \end{tabular}
\caption{{We show} the photon fluences for different time intervals around the start of the $T_{90}$ time interval. The GBM fluence is the weighted average fluence of the $85$ bursts not detected by the LAT. The LAT fluence is the weighted average fluence of the $14$ bursts detected by the LAT, {which are individually listed
in Table~\ref{GRB-Fluences}}. The LAT fluence is dominated by the three exceptionally bright bursts GRB 080916C, GRB 090902B and GRB 090926A.}
\label{PhotonFluence_Temporal}
\end{table}

Both GBM- and LAT-detected bursts show GeV-band emission long after the bulk of emission in the BATSE energy range ($50 \ \mathrm{keV}$ to $300 \ \mathrm{keV}$), which arises during the $T_{90}$ time. Whereas the high-energy photon fluence in the $T_{90}$ time interval from the GBM-detected bursts is $2.38_{-1.38}^{+1.25} \times 10^{-5} \ \gamma \ \mathrm{cm}^{-2}$, for the time interval from $0$ to $25$ times $T_{90}$ we find an average photon fluence from the GBM-detected GRBs of $19.98_{-4.77}^{+4.33} \times 10^{-5} \ \gamma \ \mathrm{cm}^{-2}$. Given that there could still be emission at later times which is just too low to be detected, the observed fluence during the $T_{90}$ time is at most $11.91_{-7.37}^{+6.
87} \%$ of the total GeV-band emission. One has to keep in mind that this is the average fluence weighted with the effective area of each burst. Since the {\it Fermi} satellite occasionally reorients for a very bright burst, our method may favor bright bursts especially for later times.

The photon fluence has also been separately calculated {and listed in Table~\ref{GRB-Fluences}} for all LAT-detected bursts using the data from $0$ to $10$ times the $T_{90}$ time interval. The results thus derived are comparable to those previously found with standard methods. The Burst Catalog \citep{2012ApJS..199...18P} can be used to compare the fluence in the high-energy region from $300 \ \mathrm{MeV}$ to $30 \ \mathrm{GeV}$ to that in the BATSE range from $50 \ \mathrm{keV}$ to $300 \ \mathrm{keV}$ obtained with the GBM. To be noted from Table \ref{GRB-Fluences} is that the bursts not detected by the LAT are fainter in the high-energy region than the LAT-detected GRBs when normalized to the same keV-band fluence.

\begin{table}[ht]
 \centering
 \begin{tabular}{c c c}
  \hline
  GRB & $F_{\mathrm{BATSE}} [\gamma \ \mathrm{cm}^{-2}]$ & $F_{[300 \ \mathrm{MeV}, 30 \ \mathrm{GeV}]} [\gamma \ \mathrm{cm}^{-2}]$ \\
  \hline
  080825C & $114.2 \pm 0.2$ & $1.81_{-1.37}^{+1.60} \times 10^{-3}$ \\
  080916C & $6.93 \pm 0.2$ & $22.27_{-2.96}^{+2.91} \times 10^{-3}$ \\
  081024B & $0.16 \pm 0.03$ & $12.93_{-6.48}^{+7.50} \times 10^{-4}$ \\
  090217 & $4.8 \pm 0.1$ & $2.12_{-0.87}^{+1.01} \times 10^{-3}$ \\
  090328 & $7.8 \pm 0.2$ & $3.59_{-1.57}^{+1.29} \times 10^{-3}$ \\
  090510 & $1.11 \pm 0.05$ & $11.99_{-1.76}^{+1.68} \times 10^{-3}$ \\
  090626 & $15.0 \pm 0.3$ & $2.19_{-1.05}^{+1.36} \times 10^{-3}$ \\
  090902B & $29.5 \pm 0.3$ & $35.87_{-3.81}^{+3.69} \times 10^{-3}$ \\
  090926A & $40.1 \pm 0.4$ & $26.64_{-3.36}^{+3.24} \times 10^{-3}$ \\
  091003 & $17.1 \pm 0.3$ & $9.99_{-5.05}^{+6.87} \times 10^{-4}$ \\
  091031 & $3.1 \pm 0.2$ & $2.55_{-1.06}^{+1.18} \times 10^{-3}$ \\
  100225A & $1.9 \pm 0.1$ & $0.95_{-0.96}^{+1.10} \times 10^{-3}$ \\
  100325A & $2.6 \pm 0.1$ &$4.39_{-4.42}^{+5.53} \times 10^{-4}$ \\
  100414A & $10.1 \pm 0.4$ & $6.83_{-3.65}^{+3.03} \times 10^{-3}$ \\
  All GBM Bursts & $3.38 \pm 0.01$ & $19.98_{-4.77}^{+4.92} \times 10^{-5}$ \\
  \hline
 \end{tabular}
\caption{Comparison of the fluences from several LAT-detected bursts and the average fluence for the $85$ 
GBM-detected bursts with the fluences in the BATSE range taken from the GRB Catalog \citep{2012ApJS..199...18P} and derived from a fit to a Band function.}
\label{GRB-Fluences}
\end{table}

\subsection{Average Flux from GRBs}

We now determine the average emission from GRBs in the energy range from $300 \ \mathrm{MeV}$ to $30 \ \mathrm{GeV}$. GRBs not detected by the LAT contribute with a fluence of $17_{-4}^{+3.7} \times 10^{-3} \ \gamma \ \mathrm{cm}^{-2}$, whereas LAT-detected GRBs account for a fluence of $120_{-7.8}^{+7.5} \times 10^{-3} \ \gamma \ \mathrm{cm}^{-2}$. Hence the total observed fluence in the range from $300 \ \mathrm{MeV}$ to $30 \ \mathrm{GeV}$ is $137_{-8.7}^{+8.3} \times 10^{-3} \ \gamma \ \mathrm{cm}^{-2}$. Since $50 \%$ of the observed fluence comes from only $2$ GRBs (GRB 090902B and GRB 090926A) the statistical error on this estimate should be roughly $0.5/\sqrt{2} \approx 35\%$. 

{To estimate the average allsky flux of GRB-produced GeV-band gamma rays one 
has to consider all cuts applied in the analysis.
First, some bursts were dropped on account of data quality as discussed in section \ref{Data Quality}. This affected $27\%$ of all available bursts giving a weight factor of $1.37$. Then, the $5^\circ$ cut on the statistical error radius of the burst location is compensated by an additional factor. The bursts excluded in this step account for $3\%$ of the overall fluence observed in the GBM energy range. This gives a factor of $1.03$, assuming that the fluences in the GBM energy range and in the LAT energy range are proportional. (We find less GeV emission than that, and therefore the true correction factor is somewhere between $1$ and $1.03$). Finally, the limitation to bursts that occurred in the LAT FoV ($\theta \le 70^\circ$) and were not too close to the galactic plane and the Blazar 3C454.3 is compensated with a factor of $3.46$. We combine the three correction factors
to derive the total efficiency factor as $4.88=3.46\cdot 1.03\cdot 1.37$. The total high-energy fluence over the entire sky should therefore be $(6.69 \pm 2.37) \times 10^{-1} \ \gamma \ \mathrm{cm}^{-2}$. The effective observation time from August 04, 2008 until July 9, 2010 was $ 5.083 \times 10^{7} \ \mathrm{s}$, and hence the total averaged flux from all bursts over the entire sky in the energy range from $300 \ \mathrm{MeV}$ to $30 \ \mathrm{GeV}$ is $(13.2 \pm 4.7) \times 10^{-9} \ \gamma \ \mathrm{cm}^{-2} \ \mathrm{s}^{-1}$. }

We have also determined the energy fluence from GRBs. 
GRBs not detected by the LAT contribute with a fluence of $1.67 \times 10^{-5} \ \mathrm{erg\,cm}^{-2}$, whereas LAT-detected GRBs account for a fluence of $1.93 \times 10^{-4} \ \mathrm{erg\,cm}^{-2}$. The average photon energies are $\sim 1$~GeV for LAT-detected GRBs and $\sim 0.6$~GeV
for GBM-detected bursts, {estimated as the ratio of the energy fluence and the photon fluence}. 
If the emission spectra were power laws, then the 
spectral indices, $s$, would be $s_{\rm LAT}\simeq 2.3$ and $s_{\rm GBM}\simeq 3$.

The average allsky energy flux is about $6.4\cdot 10^{-4}\ {\rm erg\,cm^{-2}\,yr^{-1}}$, slightly larger than the estimate of \citet{2010ApJ...722..543E}, but still compatible with it considering the uncertainties.

\section{Summary and Discussion}

In this study of high-energy emission from GRBs in the range from $300 \ \mathrm{MeV}$ to $30 \ \mathrm{GeV}$ the focus has been placed on GRBs not individually detected by the LAT and referred to as \emph{GBM detected}. We defined a sample of $85$ GRBs listed in the GBM catalog, $74$ of which can be regarded as long bursts ($T_{90} > 2 \ \mathrm{s}$) while the remaining $11$ GRBs are short ($T_{90} < 2 \ \mathrm{s}$) bursts. We find significant emission above background for the complete sample. For the long bursts the emission is clearly visible while for the short bursts alone the results are statistically inconclusive. Moreover, the GeV-band emission lasts considerably longer than the $T_{90}$ time of the keV--MeV energy range, in fact at 
least to $10$ times $T_{90}$. A similar conclusion was previously reported for individual GRBs like GRB~080916C \citep{2009arXiv0907.0714T}, GRB~081024B \citep{2010arXiv1002.3205F} or GRB~940217 \citep{1994Natur.372..652H}. Altogether only $(12\pm 7) \%$ of the total photon fluence in the range from $300 \ \mathrm{MeV}$ to $30 \ \mathrm{GeV}$ is emitted during $T_{90}$. Since this extended emission was observed when summing over many GRBs, we cannot discriminate between continous emission and a sequence of flares. Given that the number of GRBs in the sample is larger than the number of observed photons, it is also unclear whether this delayed emission is a general feature or the product of a subset of GRBs. 

The ratio of the fluence in the high-energy region to the fluence in the BATSE energy region is lower ($6\cdot 10^{-5}$) for the bursts not detected by the LAT than for LAT-detected bursts ($4.7\cdot 10^{-4}$). The estimated spectra in the 
GeV band are softer for GBM-detected bursts (with equivalent photon index $s\simeq 3$)
than for LAT-detected bursts (equivalent photon index $s\simeq 2.3$). There is no
indication that GRBs produce particle spectra with typical indices $s\simeq 2$ which are often assumed in source models of ultra-high-energy cosmic rays. Likewise,
there is no evidence for a population of GRBs that efficiently accelerate particles to high energies but individually emit too weakly in the GeV band for a
detection with {\it Fermi}-LAT.

Altogether the bursts not detected by the LAT contribute roughly $14 \%$ of all GRB-produced photons, and $8\%$ of the emitted energy, in the energy range from $300 \ \mathrm{MeV}$ to $30 \ \mathrm{GeV}$.
Finally, we find that the average allsky gamma-ray flux coming from GRBs in this energy range is $(13.16 \pm 4.65) \times 10^{-9} \ \gamma \ \mathrm{cm}^{-2} \ \mathrm{s}^{-1}$. The average allsky energy flux from GRBs in the GeV band is only $\sim 4 \%$ of the energy flux of cosmic rays above the ankle at $10^{18.6}$~eV.

\begin{acknowledgements}
We acknowledge support by the Helmholtz Alliance for Astroparticle Physics HAP
funded by the Initiative and Networking Fund of the Helmholtz Association.
\end{acknowledgements}

\bibliographystyle{aa}
\bibliography{References}

\end{document}